# Production, characterization and foamability of α-lactalbumin/glycomacropeptide supramolecular structures

Renata Silva Diniz[a], Jane Sélia dos Reis Coimbra [b,*], Alvaro Vianna Novaes de Carvalho Teixeira[c], Angélica Ribeiro da Costa[b], Igor José Boggione Santos[b], Gustavo Costa Bressan[d], Antonio Manuel da Cruz Rodrigues[a], Luiza Helena Meller da Silva[a]

[a] *Faculdade de Engenharia de Alimentos, Instituto de Tecnologia, Universidade Federal do Pará, Campus do Guamá 66000-000, Belém, PA, Brazil*
[b] *Departamento de Tecnologia de Alimentos, Universidade Federal de Viçosa, Campus Universitário s/n, Centro, 36570-900, Viçosa, MG, Brazil*
[c] *Departamento de Física, Universidade Federal de Viçosa, Campus Universitário s/n, Centro, 36570-900, Viçosa, MG, Brazil*
[d] *Departamento de Bioquímica e Biologia Molecular, Universidade Federal de Viçosa, Campus Universitário s/n, Centro, 36570-900, Viçosa, MG, Brazil*

## ABSTRACT

The study of protein interactions has generated great interest in the food industry. Therefore, research on new supramolecular structures shows promise. Supramolecular structures of the whey proteins α-lactalbumin and glycomacropeptide were produced under varying heat treatments (25 to 75 °C) and acidic conditions (pH 3.5 to 6.5). Isothermal titration calorimetry experiments showed protein interactions and demonstrated that this is an enthalpically driven process. Supramolecular protein structures in aqueous solutions were characterized by circular dichroism and intrinsic fluorescence spectroscopy. Additional photon correlation spectroscopy experiments showed that the size distribution of the structures ranged from 4 to 3545 nm among the different conditions. At higher temperatures, lower pH increased particle size. The foamability of the supramolecular protein structures was evaluated. Analysis of variance and analysis of regression for foaming properties indicated that the two-factor interactions between pH and temperature exhibited a significant effect on the volume and stability of the foam.

## 1. Introduction

Supramolecular structures are stable and well-defined chemical systems formed by molecules held together by non-covalent bonding interactions, such as electrostatic interactions, metal–ligand bonds, dispersion forces/induced, dipole-induced dipole forces, hydrophobic effects and topological bonds. The supramolecular structures are able to act in molecular recognition, catalysis and transport processes and respond to external solicitations through changes in their composition by displacing or shifting the chemical components of the solution (Lehn, 2007). This process provides materials with new functionalities with applications in different areas, especially food science.

Systems able to produce spontaneously defined and functional self-assembled architectures, such as proteins, are in the domain of supramolecular chemistry. Proteins are considered model biomaterials to produce new supramolecular structures. Therefore, controlling the protein self-assembly process can generate a wide variety of particles, such as aggregates, fibers, nanotubes and spherulites (Desfougères, Croguennec, Lechevalier, Bouhallab, & Nau, 2010).

In food systems, protein self-assembly is generally induced by heating and/or acidification. Heating usually causes partial protein unfolding, exposing hydrophobic sites hidden in the native protein. Under certain conditions, this exposure initiates self-assembly. Thus, the final products can be used in food agents for encapsulating, gelling, thickening, etc. Supramolecular structure formation is dependent on both the conformation of the proteins and the physicochemical conditions of the system (Nigen, Gaillard, Croguennec, Madec, & Bouhallab, 2010; Semenova, 2007; Van der Linden & Venema, 2007).

Among the physiochemical conditions, we emphasize: i) the pH value, which is dependent upon other physicochemical variables, e. g., ionic strength (Erabit, Flick, & Alvarez, 2014; Nicolai, Britten, & Schmitt, 2011); ii) the protein concentration, which enables interactions among proteins and consequently the formation of gels (Erabit et al., 2014); and iii) the presence of other molecules that may affect protein aggregation (Erabit et al., 2014). Therefore, understanding protein behavior as well as their structural properties in multicomponent protein systems are prerequisites for developing new sets of proteins with specific features (Nigen et al., 2010). In addition, studying the behavior of the supramolecular structure formed by different proteins can also help control the final quality of different industrial foodstuffs (Erabit et al., 2014). Therefore, better knowledge of protein behavior in solutions and their interactions with food components, including other proteins, is required when proteins are used as food ingredients at the industrial scale. In particular, heat treatment, which is widely used in the food industry, can cause protein aggregation and chemical modifications, consequently modifying the properties and behavior of components in the final product (Li, Enomoto, Ohki, Ohtomo, & Aoki, 2005; Liu & Zhong, 2013; Pinto et al., 2014).

Whey proteins are a class of proteins that can form aggregates during processing. The conditions used to form these aggregates may produce soluble materials with defined chemical and physical properties, such as surface charge, hydrophobicity, size and shape. On the other hand, these properties can act on the macroscopic properties of the final product, such as viscosity, emulsification and stability (Ryan, Zhong, & Foegeding, 2013).

Macroscopic and microscopic properties of materials containing protein assemblies are dependent on the size, structure and interactions of assemblies with the matrix material (Van der Linden & Venema, 2007). Therefore, knowing the properties, particularly the size of the aggregates, is paramount for food science research (Erabit et al., 2014), as the structure of the proteins will define their application in industry.

This study aimed to investigate the formation of supramolecular structures among the whey proteins α-lactalbumin and glycomacropeptide using heating and acidification techniques. The effect of the different experimental conditions on the conformations of the new structures was examined by applying circular dichroism and fluorescence spectroscopy. The size distribution was evaluated by photon correlation spectroscopy. Finally, we investigated the foam-forming ability of the supramolecular structures and evaluated emulsion.

## 2. Materials and methods

### 2.1. Materials

The reagents and proteins were used without further purification. The proteins α-lactalbumin (α-la) (94.9% purity) and glycomacropeptide (GMP) (95.6% purity) were provided by Davisco Foods (USA). Hydrochloric acid (HCl) was obtained from Merck (Germany). Deionized water (Milli Q, Millipore, USA; $\rho = 18.2$ MΩ·cm$^{-1}$) was used to prepare all aqueous solutions.

The individual aqueous dispersions of α-la (0.141 μM) and GMP (0.286 μM) were prepared, taking into consideration the purity of the proteins. The supramolecular structures were prepared from these suspensions by mixing defined volumes to achieve the same molar ratio defined by stoichiometric coefficient measured by isothermal titration calorimetry.

### 2.2. Differential scanning calorimetry

The thermal behavior of the proteins α-la and GMP was determined by differential scanning calorimetry (DSC) measurements (MicroCal VP-DSC, England). Aliquots of 500 μL of protein aqueous solutions at 80 μM were placed in hermetically sealed aluminum capsules. The cells were heated from 10 to 100 °C at a rate of 1 °C·min$^{-1}$. An empty capsule was used as a reference. Denaturation temperature ($T_d$) and enthalpy change ($\Delta H$) were obtained from the thermograms.

DSC was important for determining the denaturation temperature of the proteins and thus defining the heating temperatures, as the objective was to evaluate the formation of supramolecular structures at higher and lower denaturation temperatures.

### 2.3. Isothermal titration calorimetry

Electrical and chemical equipment were calibrated. The change in enthalpy of the association of GMP and α-la was monitored by an isothermal titration calorimeter (ITC) VP-ITC (MicroCal, England) at 25 °C with the aim of characterizing the interactions between these proteins.

In these experiments, a stainless steel cell filled with a 2 mL α-la aqueous solution (28 μM) was titrated with a GMP aqueous solution (250 μM). This titration was carried out by sequential injections of 300 μL of GMP titrant solution at 300 second intervals for a total of 50 injections. In addition, GMP was titrated in water and these results were subtracted from the protein titration data.

A mathematical model was used to make a nonlinear fit of the corrected heat data, resulting in a titration curve from which values of the stoichiometry ($N$), association constant ($K_a$), standard enthalpy change of interaction ($\Delta H°$) and standard Gibbs free energy of interaction ($\Delta G°$) were determined by Eq. (1). The standard entropy change ($\Delta S°$) was obtained from the relationship described by Eq. (2):

$$\Delta G° = -RT \ln K_a \qquad (1)$$

$$\Delta G° = \Delta H° - T\Delta S°. \qquad (2)$$

### 2.4. Supramolecular protein structure preparation

In order to verify the influence of pH, time, and temperature on the formation of the supramolecular protein structures, a 23 factorial experimental design with three center points was used (Table 1). pH values lower and higher than the isoelectric points of the proteins were selected to study the behavior of supramolecular structures in environments with positive and negative net charges. The temperatures used were higher and lower than the denaturation temperature of α-la to examine the effect of the denatured protein on the formation of the supramolecular structures.

Aqueous dispersions of α-la (0.141 μM) and GMP (0.286 μM) were prepared separately and homogenized under moderate agitation at room temperature for 15 min in a magnetic stirrer (TE 0851, Tecnal, Brazil). Exact volumes of each solution were mixed to reach the stoichiometric ratio (N) established by ITC analysis. To prepare 1 mL of aqueous dispersion of supramolecular structures, 254 μL of GMP dispersion (0.286 μM) was mixed with 746 μL of α-la dispersion (0.141 μM). Thus, the final concentration of GMP was 0.0726 μM and the final concentration of α-la was 0.105 μM. The protein dispersion mixture was kept under gentle agitation (TE 0851, Tecnal, Brazil) for 24 h at 4 °C (SP-500 BOD, SP Labor, Brazil).

**Table 1**
23 factorial design to evaluate the influence of pH, time and temperature variables on the protein supramolecular structure formation.

| System | pH  | Time (min) | Temperature |
|--------|-----|------------|-------------|
| 1      | 3.5 | 20         | 25          |
| 2      | 6.5 | 20         | 25          |
| 3      | 3.5 | 40         | 25          |
| 4      | 6.5 | 40         | 25          |
| 5      | 3.5 | 20         | 75          |
| 6      | 6.5 | 20         | 75          |
| 7      | 3.5 | 40         | 75          |
| 8      | 6.5 | 40         | 75          |
| 9      | 5.0 | 30         | 50          |
| 10     | 5.0 | 30         | 50          |
| 11     | 5.0 | 30         | 50          |

Afterward, the pH value of each system (HI 2221, Hanna, USA) was adjusted with HCl 0.1 mol·l$^{-1}$ and underwent its respective heat treatment. The systems were placed in an ice bath and stored at 4 °C.

### 2.5. Circular dichroism

The secondary structures of the supramolecular structures were evaluated by circular dichroism (CD). CD spectra were obtained with a JascoJ-810 spectropolarimeter (Jasco Corporation, Japan) equipped with a Peltier temperature controller (PFD 425S, Jasco, Japan) coupled to a thermostatic bath (AWC 100, Julabo, Germany). The spectra were obtained at 25 °C using a 10 mm quartz cuvette (Hellma Analytics, Germany) at a wavelength range of 190 to 260 nm. Deionized water was used as a blank.

Each spectrum was obtained by averaging ten consecutive readings. Individual dispersions of α-la (0.105 μM) and GMP (0.0726 μM) without the heat treatment and acidification were used as control. The mean residual ellipticity (MRE) was calculated using Eq. (3),

$$\text{MRE} = \frac{\Theta_{obs}}{10 n l C_p} \tag{3}$$

where $\Theta_{obs}$ is the CD in milli-degrees, $n$ is the number of amino acid residues, $l$ is the cell path length in centimeters and $C_p$ is the mole fraction.

### 2.6. Fluorescence spectroscopy

Fluorescence analyses were performed using a K2 spectrofluorometer (ISS, USA) to evaluate conformational changes in the protein supramolecular structures. Samples were analyzed in 10 mm quartz cuvettes (Hellma Analytics, Germany) at 25 °C (9001, PolyScience, USA). Fluorescence spectra were registered in the range of 290 to 450 nm. The excitation wavelength was 280 nm, which can excite the tryptophan and tyrosine residues (Lakowicz, 2006). Individual dispersions of α-la (0.105 μM) and GMP (0.0726 μM) without heat treatment and acidification were used as control.

### 2.7. Particle size distribution

The particle size of the protein supramolecular structures was determined by Photon Correlation Spectroscopy using a Zetasizer Nano S (Malvern Instrument, England) device. The instrument is equipped with an He/Ne 4 mW laser that emits a 632.8 nm wavelength, a measuring cell, avalanche photodiode detector and correlator. The samples were analyzed without dilution at (25.0 ± 0.1) °C in a rectangular polystyrene cuvette. The scattered intensity was measured under a detection angle of 173° relative to the source. Intensity autocorrelation functions were analyzed by the CONTIN algorithm.(integrated in the equipment software) to determine the size distribution. Each measurement represents an average obtained from at least ten readings and lasted approximately 60 s. Readings hindered by intensity fluctuations due to impure particles were discarded. Individual dispersions of α-la (0.105 μM) and GMP (0.0726 μM) without heat treatment and acidification were used as control.

### 2.8. Foaming ability

To evaluate the foamability of protein supramolecular structures, 20 mL of the protein suspension, in a molar ratio of 0.689:1 comprised of GMP (0.0726 μM):α-la (0.105 μM), was placed in a 50 mL beaker and homogenized (IKA Ultra Turrax T25 digital, IKA, Germany) for 72 s at 13,500 rpm. The total volume and foam volume were measured immediately after homogenization. The percentage of volume increase was calculated by the expression:

$$VI(\%) = \frac{(B - A)}{A} \times 100\% \tag{4}$$

where: $VI$ = volume increase, $A$ = protein suspension volume before agitation (mL), and $B$ = protein suspension volume after agitation (mL).

After stirring, the samples were left to stand at room temperature to assess the foam stability of the protein supramolecular structures. The variation in foam volume was measured immediately after agitation and at 5 minute intervals up to 30, 60 and 120 min. The foam stability percentage (FS) was evaluated by Eq. (5):

$$FS(\%) = \frac{V_{ft}}{V_{f0}} \times 100\% \tag{5}$$

where $V_{f0}$ = foam volume formed at time 0; $V_{ft}$ = foam volume after time $t$ (30, 60, and 120 min). The foam expansion ability ($FE$) was determined and calculated according to Eq. (6):

$$FE(\%) = \frac{V_{f0}}{V_{il}} \times 100\% \tag{6}$$

where $V_{f0}$ = foam volume formed; $V_{il}$ = liquid initial volume.

Individual dispersions of α-la (0.105 μM) and GMP (0.0726 μM) without the heat treatment and acidification were used as controls.

### 2.9. Emulsifying properties

To determine the emulsifying properties of the supramolecular protein structures, the emulsifying activity index (EAI) and emulsion stability index (ESI) were determined according to Pearce and Kinsella (1978) and Guo and Mu (2011), with some modifications.

The mixture containing 30 mL of protein dispersions, in a molar ratio of 0.689:1 comprised of GMP (0.0726 μM):α-la (0.105 μM) and 10 mL of soy oil was kept under agitation (IKA Ultra Turrax T25 digital, IKA, Germany) for 1 min at 24,000 rpm. Aliquots of the emulsion (50 μL) were pipetted from the bottom of the container at 0 and 10 min after homogenization and diluted 100-fold using 0.1% sodium dodecyl sulfate aqueous solution (SDS). The absorbance of the diluted solution was measured at 500 nm (Cary 50, Varian, Australia). Absorbance values, measured immediately after agitation stopped ($A_0$) and 10 min ($A_{10}$) after emulsion formation, were used to calculate EAI and ESI properties.

The emulsifying activity index (m2·g$^{-1}$) and emulsion stability index (min) were calculated according to Guo and Mu (2011), with Eqs. (7) and (8).

$$EAI = \frac{2 \times 2.303 \times A_0 \times D}{c \times l \times (1 - \varphi) \times 1000} \tag{7}$$

$$ESI = \frac{A_0}{A_0 - A_{10}} \times t \tag{8}$$

where $c$ is the initial protein concentration, $\varphi$ is the volume fraction of oil used in the emulsion (0.25), $D$ is the dilution factor used (100), $t$ is 10 min and $A_0$ and $A_{10}$ are the diluted emulsion absorbances at times (0 and 10) in min.

Individual dispersions of α-la (0.105 μM) and GMP (0.0726 μM) without heat treatment and acidification were used as control.

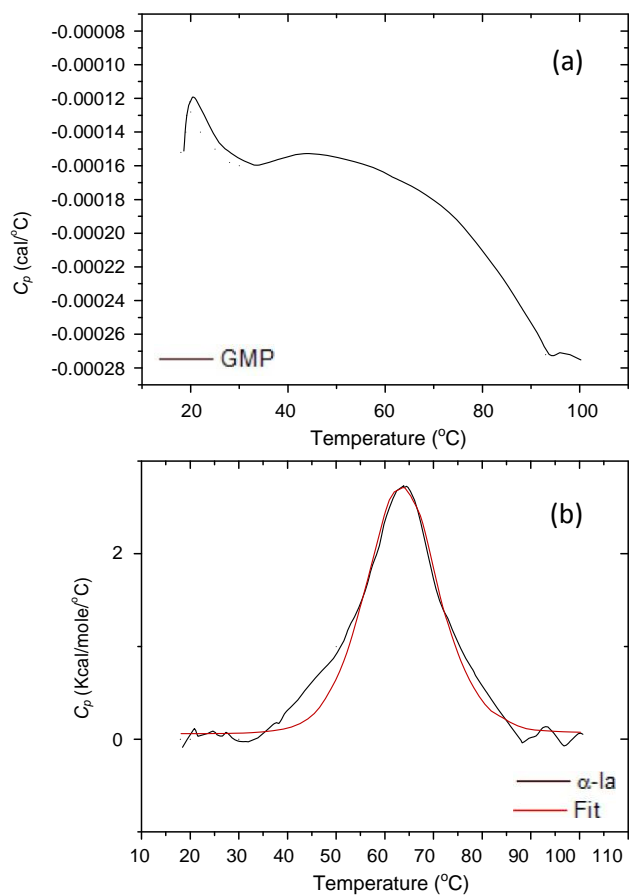

**Fig. 1.** Differential scanning calorimetry of proteins (a) GMP 80 μM and (b) α-la 80 μM; red line represents curve fit. (For interpretation of the references to color in this figure legend, the reader is referred to the web version of this article.)

### 2.10. Statistical analysis

The experimental data obtained for the techno-functional properties for all supramolecular structures formed (Table 1) were statistically analyzed using regression analysis (5% probability).

## 3. Results and discussion

### 3.1. Differential scanning calorimetry

Fig. 1 shows the thermal behavior of the proteins with the objective of evaluating the formation of supramolecular structures at higher and lower denaturation temperatures.

There was no well-defined curve for GMP because it has no defined tertiary structure, being characterized as an intrinsically unstructured peptide with no sulfide bonds (Martinez, Sanchez, Patino, & Piloso, 2009; Neelima, Sharma, Rajput, & Mann, 2013).

The thermogram obtained for α-la is characterized by an endothermic process where the peak curve for Cp vs. temperature was 62.71 °C, corresponding to the denaturation temperature of this protein. De Souza, Bai, Gonçalves, and Bastos (2009) found a higher value for α-la denaturation temperature (65.04 °C) in a calorimetric study on whey protein isolate. In a study of thermal properties of protein aggregates, Ju, Hettiarachchy, and Kilara (1999) observed that α-la denaturation occurs at a temperature range of 59 to 71 °C and the endothermic peak occurs at 67.1 °C. These differences, although small, may result from changes in factors such as protein source, pH and ionic strength. The value of the enthalpy change ($\Delta H$) of the process was $46.18 \pm 0.25$ kcal·mol$^{-1}$.

### 3.2. Isothermal titration calorimetry

The effect of associating GMP with α-la was assessed by isothermal titration calorimetry. The values of ΔH represent the enthalpy energy absorbed or released when GMP interacts with α-la. The magnitude of $\Delta H$ reflects the contributions of four other processes:

(i) association/dissociation of ions with charged groups on the proteins ($\Delta_{int}H_i$);
(ii) association of GMP with α-la ($\Delta_{int}H_{GMP-\alpha-la}$);
(iii) changes in the solvation of the proteins ($\Delta_{int}H_{sol}$);
(iv) rearrangement of the protein molecular structure ($\Delta_{int}H_{conf.rear}$).

Thus:

$$\Delta H° = \Delta_{int}H_i + \Delta_{int}H_{GMP-\alpha-la} + \Delta_{int}H_{sol} + \Delta_{int}H_{conf.rear}. \quad (9)$$

The first contribution, $\Delta_{int}H_i$, is associated with the interaction of ions in solution with charged groups in the protein molecules, which is an exothermic process. The second contribution, $\Delta_{int}H_{GMP-\alpha-la}$, refers to the interaction between GMP-α-la. These interactions release energy, contributing to an exothermic $\Delta H°$. The third contribution, $\Delta_{int}H_{sol}$, refers to the desolvation of the proteins, which is an endothermic process since it requires the interactions of the water-GMP, GMP-ion, ion-α-la, and water-α-la to be broken. The fourth contribution, $\Delta_{int}H_{conf.rear}$, is associated with protein conformational change, which is endothermic as the intramolecular interactions are broken.

The titration of GMP solution in α-la solution, already subtracted from the blank, produced the curve shown in Fig. 2.

Values of $\Delta_{int}H$ are initially exothermic and become increasingly more negative due to the contributions of endothermic processes in accordance to Eq. (9): $\Delta_{int}H_{sol}$ and $\Delta_{int}H_{conf.rear}$. In addition to the disruption of the interactions in the process $\Delta_{int}H_{sol}$ above, the protein–water and protein–ion interactions are also broken. These interactions are quite intense, as they are characterized by hydrogen bonds and electrostatic attractions, respectively. Furthermore, in the interaction between proteins GMP-α-la, the proteins change their conformation in such a way that the hydrophilic groups are exposed and free to interact with the solution, as seen in the fluorimetry results. Increasing the GMP-α-la molar ratio increases the contribution of the endothermic processes until these values are similar to those of the enthalpy of the exothermic processes. The change in enthalpy then becomes approximately zero, as shown in Fig. 2.

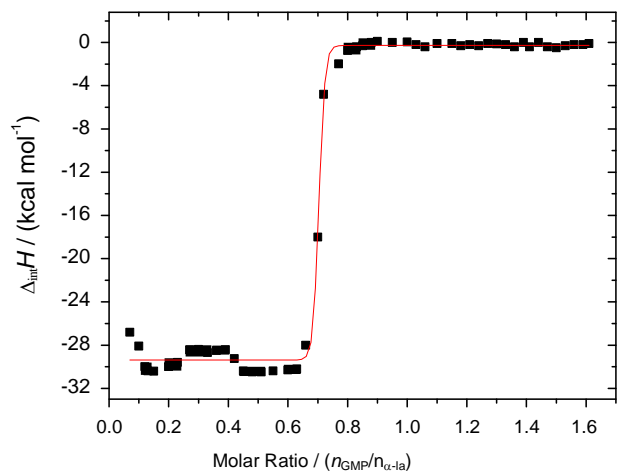

**Fig. 2.** Isothermal titration calorimetry of GMP and α-la protein.

As the population of α-la molecules becomes saturated with GMP, the thermal signal gradually decreases. The curve of the heat integration according to the molar ratio between the reactants was adjusted and the thermodynamic interaction parameters, such as stoichiometric coefficient ($N$), the equilibrium constant ($K$), the standard enthalpy change of interaction ($\Delta H°$), standard free energy change of interaction ($\Delta G°$) and the standard entropic change ($\Delta S°$) were determined using the MicroCal VP-DSC software. Data are shown in Table 2. The stoichiometric coefficient obtained was 0.689, thus the supramolecular structures were produced by mixing dispersions of GMP (0.286 μM) and α-la (0.141 μM) in a molar ratio of 0.689:1 (GMP:α-la). Under these conditions, we observed that α-la and GMP have strong interaction due to the value of $\Delta H°$ being $-2.963 \times 10^4$ cal·mol$^{-1}$.

Entropic contribution balancing occurs by two processes. The first is the conformational change of the proteins. This conformation change reduces the degree of translational freedom of the species charged and therefore decreases the entropy of the system. The second process is related to the GMP-α-la interaction. For GMP to interact with α-la, both proteins need to release the solvation molecules ($H_2O$ and ions). This process releases the molecules into the solution, increasing the configurational entropy of the system. In this case, the entropy is lower than zero ($-61.40$ cal/mol·K), thus the first factor prevails over the second factor. Therefore, we conclude that interaction occurs between the proteins α-la and GMP and the process is enthalpically driven.

### 3.3. Circular dichroism

Fig. 3 shows the CD spectra of the native forms of α-la and GMP and the protein supramolecular structures. The CD spectrum of native α-la is typical of well-defined α-helix-rich proteins because of its negative peaks at approximately 222 nm and 208 nm (Fig. 3a–c). Native GMP, on the other hand, shows a minimum peak around 198 nm, thus characterizing its secondary structure as a typical random coil.

Along with the calorimetry results, these CD spectra obtained here show that the supramolecular structures are indeed formed in solution. The CD spectra obtained for all supramolecular complexes maintained the overall characteristics expected for an α-helix-rich protein structure, although the negative peak around 208 nm observed for native α-la was shifted to around 205 nm, possibly due to the contribution of GMP in the system. We noted an increased negative signal intensity around 205–208 nm in all supramolecular structures analyzed, suggesting that new regular secondary structures may have formed during supramolecular structure assembly.

The treatments confirmed that increased temperature did not cause substantial changes in the overall secondary structure pattern of the supramolecular complexes at pH 3.5. On the other hand, this increase in temperature leads to a decrease in the negative molar ellipticity at pH 6.5 (Fig. 3a). This same behavior was observed by Faizullin, Konnova, Haertle, and Zuev (2013), studying β-casein at similar increasing temperature conditions. At 75 °C, the system at pH 3.5 had more intense negative peaks than the system at pH 6.5, similar to results observed by Naqvi, Ahmad, Khan, and Saleemuddin (2013) studying β-lactoglobulin, which displayed higher negative molar ellipticity at pH 2.0 compared to pH 7.0 and 9.0. These findings indicate that in the systems analyzed, lower pH might favor the formation of secondary structures at increased temperatures.

**Table 2**
Thermodynamic parameters of the α-la and GMP interaction, at 25 °C.

| | |
|---|---|
| $N$ | 0.689 |
| $K$ (M$^{-1}$) | $2.03 \times 10^8$ |
| $\Delta H°$ (kcal/mol) | $-29.63$ |
| $\Delta S°$ (cal/mol·K) | $-61.40$ |
| $T\Delta S$ (kcal/mol) | $-18.31$ |
| $\Delta G°$ (kcal/mol) | $-11.32$ |

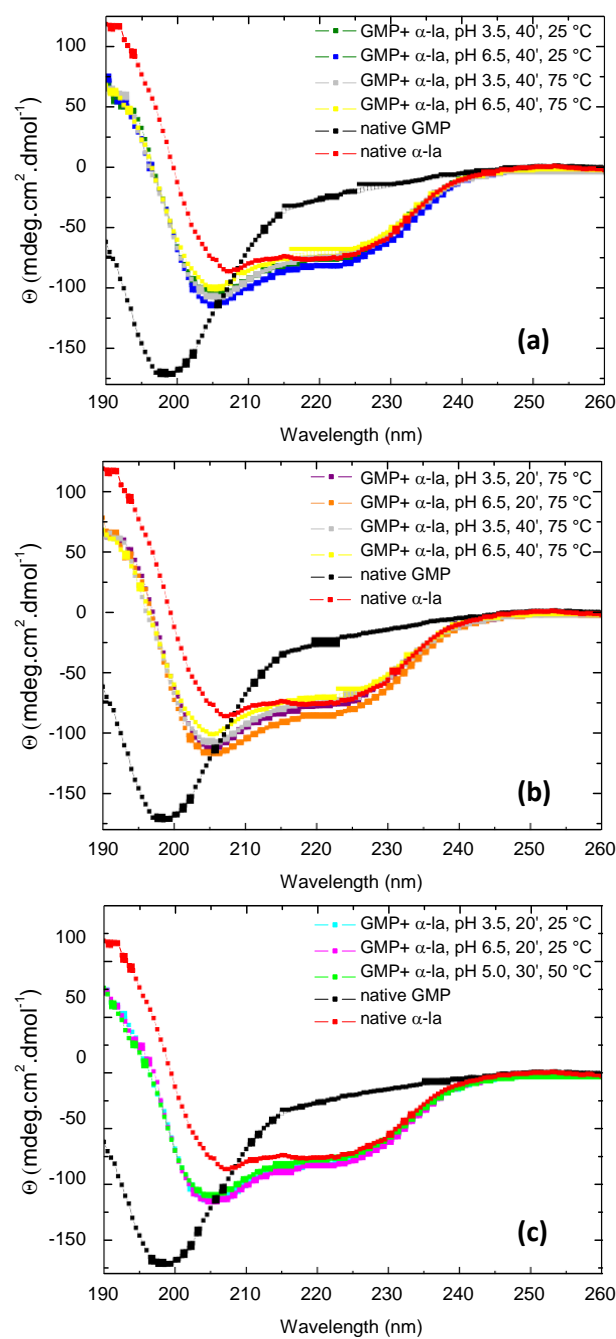

**Fig. 3.** CD spectra of the native proteins α-la (red) and GMP (black) and supramolecular structures in different systems. (For interpretation of the references to color in this figure legend, the reader is referred to the web version of this article.)

Regarding time exposed to heat, we observed no prominent differences in the CD spectra obtained for supramolecular structures at pH 3.5. However, at pH 6.5, longer heat exposure at 75 °C decreased the negative molar ellipticity of the system, indicating possible loss of regular secondary structures. These results differed from those of Moro, Báez, Busti, Ballerini, and Delorenzi (2011), who studied β-lactoglobulin at 85 °C. They observed that the negative CD spectra peaks increased in intensity with longer heating times. Bobály et al. (2014), on the other hand, observed increased molar ellipticity values in CD spectra for lysozymes in acidic conditions, similar to what we observed in our systems at pH 3.5 and 6.5, heated to 75 °C for 40 min (Fig. 3b). Together, these data show that the acidic condition tested here seems to contribute to maintain the overall content of the secondary structures with longer heat exposure.

For the supramolecular structure prepared at pH 5.0 at 50 °C for 30 min there was a short decrease in the intensity of the negative band compared to the systems at 25 °C, indicating that the secondary structure of this system was altered with the change in pH and increased temperature. Zhang and Zhong (2012) heated bovine serum albumin and verified that increased temperature reduced the intensity of the negative peaks in CD spectra.

### 3.4. Fluorescence spectroscopy

GMP does not present intrinsic fluorescence emission, as its primary sequence does not contain fluorescent amino acid residues such as tyrosine, phenylalanine, and tryptophan (Neelima et al., 2013). On the other hand, α-la has four tryptophan residues in its primary structure at positions 26, 60, 104 and 118 and four tyrosine residues at positions 18, 36, 50 and 103 (Sgarbieri, 2005). Thus, possible changes in the chromophore environment during the supramolecular structure formation could be monitored through intrinsic fluorescence spectroscopy.

While GMP did not exhibit the expected intrinsic fluorescence, the emission spectrum of native α-la was characterized by a maximum emission at 338 nm, similar to Zhang et al. (2014). Intrinsic fluorescence spectra of the native proteins α-la and GMP and of the supramolecular structures excited at 280 nm are shown in Fig. 4.

In general, compared to the emission spectrum of native α-la, the maximum emission peak of all systems analyzed shifted to shorter wavelengths (approximately 322 nm). In addition, an increase in peak emission intensity was observed. These findings indicate that the surrounding chemical environment of the chromophores became more apolar (Jindal & Naeem, 2013; Taheri-Kafrani, Asgari-Mobarakeh, Bordba, & Haertlé, 2010; Taheri-Kafrani, Bordbar, Mousavi, & Haertlé, 2008; Zhang, Qi, Zheng, Li, & Liu, 2009), suggesting that during the interaction between α-la and GMP molecules, the chromophores remain buried deeper in the core structures composed of these molecules (Fig. 4a and b). Furthermore, we found that the systems formed at pH 6.5 presented higher fluorescence intensity (Fig. 4a–c), indicating that this pH favors the interaction between the proteins α-la and GMP, since the decrease in polarity may result from the chromophores being further hidden inside the supramolecular structures. These results, together with those obtained by calorimetry and circular dichroism, confirmed that the association between these proteins occurs in the conditions analyzed.

Comparing the systems at pH 3.5 with those at pH 6.5, increased temperature reduced fluorescence intensity and slightly shifted the maximum fluorescence emission intensity to higher wavelengths, indicating that the chromophore neighborhood became more polar and more exposed to the solvent. Zhang and Zhong (2012) also observed that heat decreased bovine serum albumin fluorescence intensity. These findings show that the tridimensional features of the supramolecular structures formed might be different among the temperatures analyzed, even though the overall content of secondary structures remained constant (Figs. 3a and 4a).

Regarding the heat exposure time, the supramolecular structures maintained at pH 6.5 presented higher fluorescence intensity when incubated for 40 min at 75 °C, although no significant difference was observed among exposure times for the systems at pH 3.5 (Fig. 4b). Moreover, the supramolecular structures formed at pH 6.5 showed more intense fluorescence emission peak than those formed at pH 3.5 (Fig. 4a and b). This observation suggests that the less acidic pH solution provides a more suitable environment for the structures when they are exposed to higher temperatures for a longer time. Naqvi et al. (2013) studied the intrinsic fluorescence of β-lactoglobulin and observed that at pH 9.0, the protein showed higher fluorescence intensity than at pH 7.0, in accordance with our results.

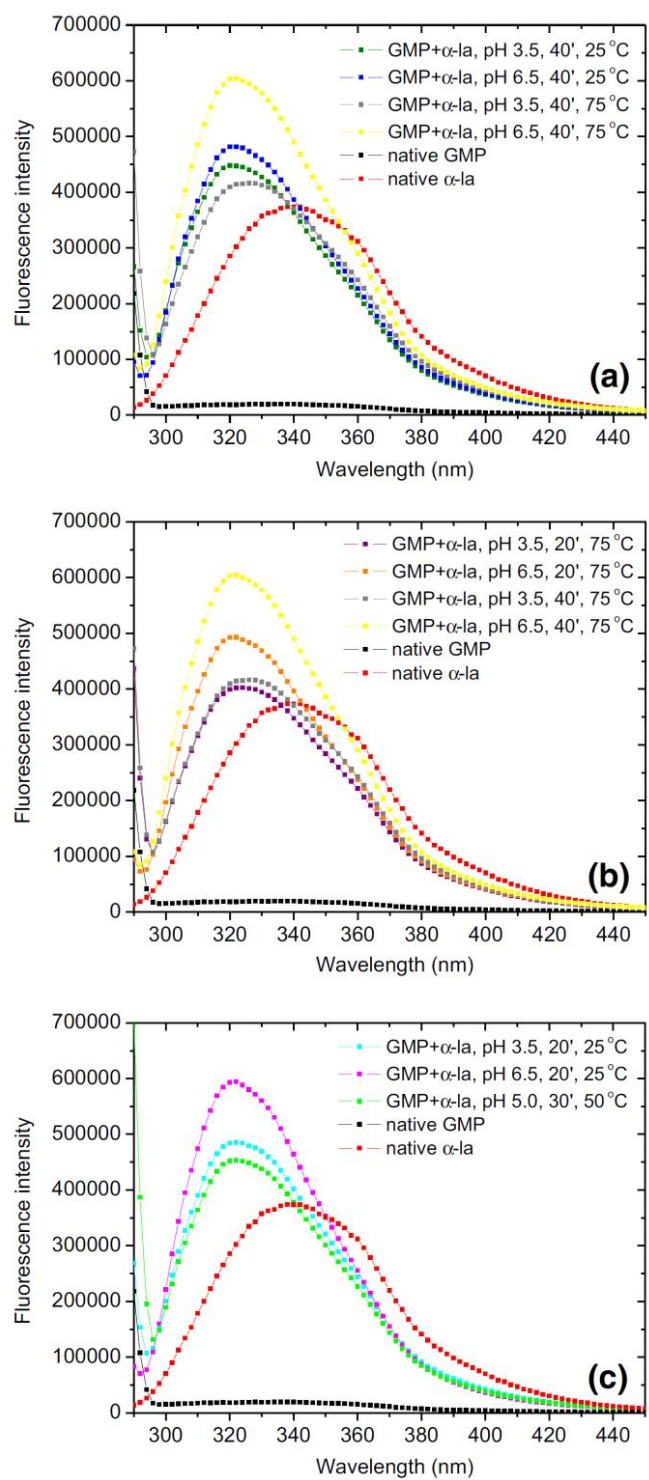

**Fig. 4.** Emission spectra of chromophores of the native proteins α-la (red) and GMP (black), and of the protein supramolecular structures in different systems. (For interpretation of the references to color in this figure legend, the reader is referred to the web version of this article.)

Fig. 4c shows that systems treated at 50 °C for 30 min with pH 5.0 showed less intense emission spectra than systems with pH 6.5 and 3.5 treated at 25 °C for 20 min. This indicated that the protein underwent conformational changes around the chromophores due to heating, turning its neighborhood more polar when compared to 25 °C (Zhang et al., 2014).

### 3.5. Particle size distribution

The size of the protein supramolecular structures was determined by analysis in a Zetasizer device. The results of samples and controls (individual preparations of α-la and GMP without acidification and without heat treatment) are shown in Table 3. The main populations defined here refer to those with the highest percentage in the particle volume distribution.

α-la is a compact globular protein, which has the dimensions of $2.5 \times 3.7 \times 3.2$ nm in solution (Fox & McSweeney, 1998). After particle size analysis, we found that α-la had a particle population of 2 nm in diameter, indicating no change in structure during homogenization. On the other hand, the GMP solution had three different populations with sizes larger than the protein particle. The value observed for the GMP main population (350 nm) may be caused by the aggregate formation of these proteins, indicating that the aqueous medium (where the solution has been prepared) is not the ideal vehicle for this protein. Furthermore, it is difficult to analyze size distributions for pure proteins, since larger impurities, although fewer in number, mask the true size of the protein. In systems 2, 4, 5, 7, 8 and 10, particle populations with larger sizes (1178–5581 nm) were observed. However, these populations were less common, which suggests that these values may be related to the presence of impurities in the samples.

The protein supramolecular structures formed at 25 °C for 20 and 40 min at pH 6.5 had a main population with a size of about 265 nm. On the other hand, the structures formed at 25 °C for 20 and 40 min at pH 3.5 had approximate sizes of 4 nm for the main population, indicating that the more acidic pH (3.5) at 25 °C does not cause the association between the proteins.

The heat treatment at 75 °C in samples with pH 3.5 originated a main population with particle sizes of 1311 and 3545 nm for the 20 and 40 minute treatment times. This result suggests that a longer heating time induces the formation of larger aggregates, as reported by Ryan et al. (2013). The main populations of systems at pH 6.5 formed by heating at 75 °C for 20 and 40 min had sizes of 270 and 238 nm. These values are similar to those observed for systems with the same pH at 25 °C, indicating that at pH 6.5 the heat treatment does not influence the size of supramolecular structures. For structures formed at 75 °C, more acidic pH (3.5) causes the formation of larger particles than those formed by pH 6.5. Phan-Xuan et al. (2011) obtained similar results for β-lactoglobulin microgels. They observed that when pH increased, β-lactoglobulin microgel particle size decreased.

Furthermore, we observed that for systems at pH 3.5, heat treatment at 25 °C did not cause protein aggregation, while heat treatment at 75 °C formed aggregate micrometers in size. This result is in accordance with Le, Saveyn, Hoa, and der Meeren (2008), who verified the larger size of casein and whey protein aggregates with higher temperature. Dybowska (2011) also observed larger whey protein concentrate particles after heat treatment. The systems formed by heat treatment at 50 °C for 30 min and pH 5.0 had aggregates with main population particle sizes between 1124 and 1597 nm. This indicates that higher temperatures enable larger aggregates to form, since the structures formed at 25 °C originated particles of at most 266 nm.

### 3.6. Foaming ability

To evaluate possible improvements in functional properties of supramolecular structures, foamability was tested at 30, 60 and 120 min. Some systems did not remain stable at 60 and 120 min, therefore only results for foamability (VI, FS, FE) at 30 min were examined by analysis of variance (ANOVA) and regression analysis. The individual effect of each factor is measured by the mean coefficient of the main effect. The effect of two or more factors is expressed as the coefficient in two- factor or three-factor interactions. ANOVA showed that pH, time and temperature levels had no significant (p N 0.05) effect on the properties studied. Only two-factor interactions (pH and temperature) exhibited a significant effect on VI and FS properties, however, indicating that such factors do not work independently and the effect of each factor depends on the level of other factors (Table 4, Fig. 5). On the other hand, for FE, the main effects or interactions were not significant.

**Table 3**
Particle size distribution of protein supramolecular structures.

|  | Mean size diameter (nm) |  | % Volume dstribution |
|---|---|---|---|
| α-la | Peak 1 | 2 | 100 |
| GMP | Peak 1 | 350 | 77.4 |
|  | Peak 2 | 64 | 14.2 |
|  | Peak 3 | 5227 | 8.4 |
| System 1 | Peak 1 | 4 | 98.9 |
|  | Peak 2 | 13 | 0.6 |
|  | Peak 3 | 27 | 0.1 |
| System 2 | Peak 1 | 265 | 83.8 |
|  | Peak 2 | 5151 | 8.6 |
|  | Peak 3 | 60 | 7.6 |
| System 3 | Peak 1 | 4 | 97.9 |
|  | Peak 2 | 11 | 1.9 |
| System 4 | Peak 1 | 266 | 82.2 |
|  | Peak 2 | 4871 | 10.7 |
|  | Peak 3 | 54 | 7.1 |
| System 5 | Peak 1 | 1311 | 98.9 |
|  | Peak 2 | 5581 | 1.0 |
|  | Peak 3 | 205 | 0.1 |
| System 6 | Peak 1 | 270 | 88.6 |
|  | Peak 2 | 60 | 11.4 |
| System 7 | Peak 1 | 3545 | 80.6 |
|  | Peak 2 | 1178 | 18.8 |
|  | Peak 3 | 439 | 0.7 |
| System 8 | Peak 1 | 70 | 10.8 |
|  | Peak 2 | 238 | 77.3 |
|  | Peak 3 | 5262 | 11.9 |
| System 9 | Peak 1 | 1397 | 65.1 |
|  | Peak 2 | 5300 | 34.9 |
| System 10 | Peak 1 | 1597 | 99.8 |
|  | Peak 2 | 5536 | 0.2 |
| System 11 | Peak 1 | 1124 | 100.0 |

**Table 4**
Summary of factory analysis of pH, time and, temperature on volume increase (VI) and foam stability (FS).

| Term | Coefficient ($\beta_i$) for VI | P | Coefficient ($\beta_i$) for FS | P |
|---|---|---|---|---|
| Constant | 198.864 | 0.000 | 368.864 | 0.000 |
| Main effect |  |  |  |  |
| A – pH | −18.75 | 0.514 | −39.375 | 0.428 |
| B – time | a | a | a | a |
| C – temperature | −1.958 | 0.514 | −2.920 | 0.16 |
| 2-Way interactions |  |  |  |  |
| AB | a | a | a | a |
| AC | 0.417 | 0.011 | 0.692 | 0.019 |
| BC | a | a | a | a |

[a] Not statistically significant ($p > 0.05$).

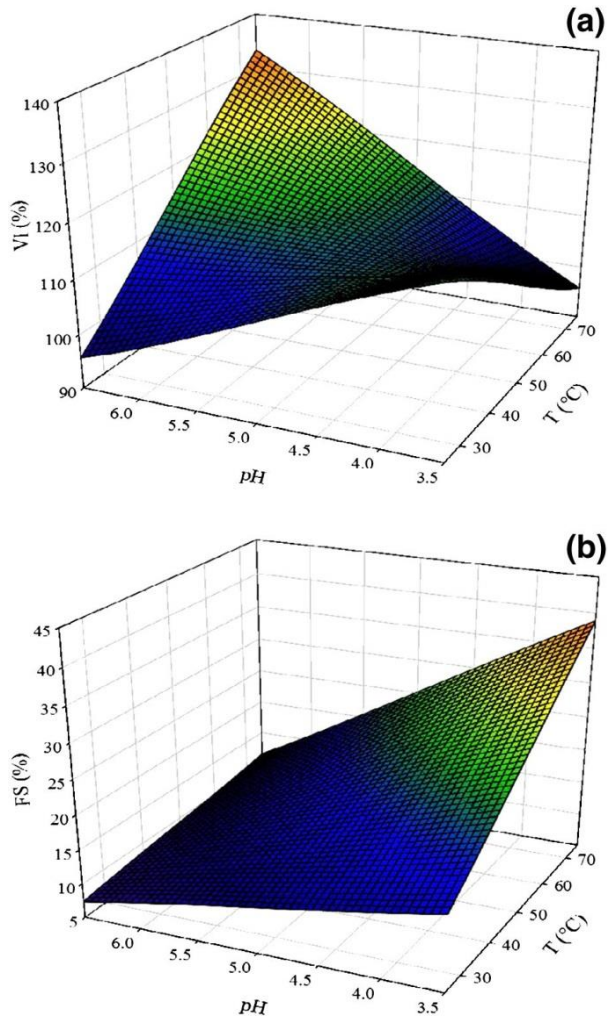

**Fig. 5.** (a) Three-dimensional response surface for volume increase (VI) as a function of pH and temperature for supramolecular structures. (b) Three-dimensional response surface for foam stability (FS) as a function of pH and temperature for supramolecular structures.

The supramolecular structure system that had the highest ability to increase foam volume was formed at high temperature and high pH (pH 6.5, 75 °C), as noted by Kuropatwa, Tolkach, and Kulozik (2009). These authors studied the impact of pH on the foamability of whey protein–egg protein complex at high temperature and found that increased pH leads to an increase in foam volume. CD spectra of these supramolecular structures show that this treatment induced changes in the secondary structure of proteins with a lower ellipticity value compared to other treatments. Furthermore, these supramolecular structures had higher fluorescence intensity, suggesting that the neighborhood of the chromophores became more apolar and foamability was enhanced by increased protein hydrophobicity. These conformational changes lead to surface- particles capable of strongly packing at the air/water interface and forming a viscoelastic network. These results are in agreement with those reported by Kim, Cornec, and Narsimhan (2005), who studied the functional properties of β-lactoglobulin at different pH and heated to 80 °C. They found that the protein heated at pH 5.5 for 15 min showed the best foamability.

FS properties improved when temperature increased and pH decreased. This may be explained by pH 3.5 being close to the isoelectric point of the proteins, which have low electrostatic interactions as well as hydrophobic interactions at the air–water interface which may be responsible for foam stabilization (Kuropatwa et al., 2009). Furthermore, higher temperature positively influenced foam stability due to partial protein unfolding, which in turn increases hydrophobicity, resulting in increased protein–protein interactions (Moro et al., 2011; Nicolai et al., 2011; Nicorescu et al., 2010). Although pH 3.5 is closer to the isoelectric point, the supramolecular structures formed at pH 6.5 were better able to increase foam volume at 75 °C, because the structures formed at a higher pH presented smaller particle size. One of the most important factors for foam formation is the protein adsorption rate, which depends on protein concentration, protein size, protein structure and solution conditions (Martin, Grolle, Bos, Cohen Stuart, & van Vliet, 2002; Moro et al., 2011). Thus, disordered, smaller and more flexible proteins are better surface agents than ordered, larger and rigid ones (Moro et al., 2011). The GMP solution slightly influenced volume increase and foam expansion, suggesting that α-la is the protein responsible for foamability (data not shown).

### 3.7. Emulsifying properties

To evaluate the emulsifying properties of the protein supramolecular structures, the emulsifying activity index (EAI) and emulsion stability index (ESI) were calculated by the turbidimetric method. The results for EAI and ESI were examined by analysis of variance (ANOVA) and regression analysis. ANOVA showed that pH, time and temperature levels had no significant (p N 0.05) effect on the properties studied.

### 4. Conclusions

Supramolecular structures can be formed by heating and acidifying aqueous solutions of α-la and GMP. CD results showed changes in the secondary structure of protein supramolecular structures compared to the native proteins. The fluorescence analysis indicated that the supramolecular structures have a more hydrophobic core to their chromophores. The results of the particle size distribution show that the proteins are associated in different ways, depending on the heat treatment condition and association. Particle diameter ranged from nanometers to micrometers. At high temperature and high pH, the foam volume of the supramolecular structures increased. Furthermore, pH, time and temperature levels had no significant effect on emulsifying properties. The formation of supramolecular structures offers new opportunities to modify the functional properties of food. Heating and acidifying conditions can be optimized to control the properties of the proteins and to improve their performance in an increasing number of applications. In addition, the results allowed better knowledge of the factors that can help predict the behavior of protein associations in complex systems.

### Acknowledgments

This study was supported by CAPES, CNPq, and FAPEMIG. The authors thank the Brazilian Biosciences National Laboratory – Spectroscopy and Calorimetry Laboratory – and Núcleo de Microscopia e Microanálise (Universidade Federal de Viçosa). The authors thank prof. Dr. Nilda de Fátima Ferreira Soares and MSc. Éverton de Almeida Alves Barbosa for their support for this work.